\documentclass[preprints,article,accept,moreauthors,pdftex]{Definitions/mdpi}

\firstpage{1} 
\pubvolume{1}
\issuenum{1}
\articlenumber{0}
\pubyear{2021}
\copyrightyear{2020}
\datereceived{} 
\dateaccepted{} 
\datepublished{} 
\hreflink{https://doi.org/} 

\Title{How Can Autonomous Vehicles Convey Emotions to Pedestrians? A Review of Emotionally Expressive Non-Humanoid Robots}

\TitleCitation{How Can Autonomous Vehicles Convey Emotions to Pedestrians? A Review of Emotionally Expressive Non-Humanoid Robots}

\Author{Yiyuan Wang *, Luke Hespanhol and Martin Tomitsch}

\AuthorNames{Yiyuan Wang, Luke Hespanhol and Martin Tomitsch}

\AuthorCitation{Wang, Y.; Hespanhol, L.; Tomitsch, M.}

\address[1]{%
Design Lab, Sydney School of Architecture, Design and Planning, University of Sydney, 2006 Sydney, Australia}

\corres{Correspondence: ywan8818@uni.sydney.edu.au}

\abstract{In recent years, researchers and manufacturers have started to investigate ways to enable autonomous vehicles (AVs) to interact with nearby pedestrians in compensation for the absence of human drivers. The majority of these efforts focuses on external human-machine interfaces (eHMIs), using different modalities such as light patterns or on-road projections to communicate the AV's intent and awareness. In this paper, we investigate the potential role of affective interfaces to convey emotions via eHMIs. To date, little is known about the role that affective interfaces can play in supporting AV-pedestrian interaction. However, emotions have been employed in many smaller social robots from domestic companions to outdoor aerial robots in the form of drones. To develop a foundation for affective AV-pedestrian interfaces, we reviewed the emotional expressions of non-humanoid robots in 25 articles published between 2011 and 2021. Based on findings from the review, we present a set of considerations for designing affective AV-pedestrian interfaces and highlight avenues for investigating these opportunities in future studies.}

\keyword{Emotional expression, non-humanoid robots, autonomous vehicles, pedestrians, human-machine interfaces, human-robot interaction} 

\RequirePackage[normalem]{ulem} 
\RequirePackage{color}\definecolor{RED}{rgb}{1,0,0}\definecolor{BLUE}{rgb}{0,0,1} 
\providecommand{\DIFadd}[1]{{\protect\color{blue}\uwave{#1}}} 
\providecommand{\DIFaddbegin}{} 
\providecommand{\DIFaddend}{} 
\providecommand{\DIFaddbeginFL}{} 
\providecommand{\DIFaddendFL}{} 
\providecommand{\DIFdelbeginFL}{} 
\providecommand{\DIFdelendFL}{} 

\begin{document}

\section{Introduction}
The introduction of autonomy in vehicles promises to increase the level of convenience and comfort for riders \cite{litman2017autonomous}. However, the absence of human drivers in fully autonomous systems induces an interaction void around what used to be traditional communication strategies between drivers and other road users, such as eye contact and body gestures \cite{rasouli2019autonomous}. The responsibility for communicating internal states and intentions in typically short, dynamic traffic scenarios is to a large extent, if not entirely, delegated from drivers to the self-moving vehicles themselves.

Researchers and manufacturers have been dedicated to developing additional channels to assist autonomous vehicles (AVs) in conveying their intent and awareness to surrounding road users, especially to pedestrians, who are considered as one of the most vulnerable yet frequent interaction subjects \cite{rasouli2019autonomous}. Existing means that aim to support a safe and intuitive AV-pedestrian interaction are various, ranging from display technologies such as LED lighting patterns \cite{lagstrom2016avip,Mahadevan2018communicating} and on-road projections \cite{Nguyen2019designing,prattico2021comparing}, to borrowing anthropomorphic features such as moving eyes that follow the pedestrian's position \cite{chang2017eyes,locken2019automated} or displaying a smiling expression on the front of AVs \cite{locken2019automated,prattico2021comparing}.

While the existing solutions have been validated to compensate for the lack of driver-pedestrian communication in different ways, emotions, a vital dimension in human-human interaction \cite{cauchard2016emotion}, have thus far been mostly disregarded in this growing area of research. Different from pragmatic channels, emotions have a unique role in affecting perception, empathy, decision-making, and social interactions \cite{picard2003affective}. In fact, imbuing emotions into social robots is not rare in human-robot interaction (HRI). Prior work describes the ability to convey emotions as one of the indicators of socially interactive robots \cite{fong2003survey}. Previously designed robots that are capable of expressing emotions vary in functionalities, appearances, as well as how they articulate emotions in interactional contexts. We reviewed a sub-domain of these robots --- emotionally expressive robots that have a non-humanoid form --- aiming to translate the design paradigms for their emotional expressions into cues that AVs may employ in the affective dimension, as AVs are in essence self-moving, non-humanoid social robots.

The contribution of this paper is twofold. First, we systematically review 25 articles focusing on designing and evaluating emotional expressions for non-humanoid robots in the past ten years. We summarise emotion models, output modalities, evaluation measures, as well as how users perceived the emotional expressions in these studies. Second, based on the findings from the review, we propose a set of considerations for designing affective AV-pedestrian interfaces by adding emotions as an additional communication dimension. Our findings contribute to enhancing AV-pedestrian interaction and increasing social acceptance for the deployment of AVs in urban environments.

\section{Background}
\subsection{Emotion in Social Robotics}
Social robots have been utilised in many domains including healthcare, education, domestic environments, and public spaces \cite{leite2013social}. To bring the interactions with humans onto a more natural and engaging level \cite{bretan2015emotionally,gacsi2016humans,ritschel2019personalized}, many of these robots have been integrated with the capability of displaying emotions. Eyssel et al. \cite{eyssel2010anthropomorphic} showed that a robot expressing emotional states could make people feel closer to the robot, perceive it as having anthropomorphic traits and intentionality, and experience a more pleasant HRI process. Besides facilitating empathic inferences about internal states and intentions \cite{loffler2018multimodal}, emotions expressed by robots can subsequently elicit affective responses from humans \cite{boccanfuso2015autonomously,Herdel2021drone,Whittaker2021designing} and impact the immediate environment where the interaction is taking place \cite{harris2011exploring,Hoggenmueller2020emotional}. With the increased perceived sociability \cite{Herdel2021drone}, emotionally expressive robots can effectively establish closer bonds with humans and, more importantly, increase public trust and acceptance \cite{cauchard2016emotion,Hoggenmueller2020emotional,Tennent2018Character,Whittaker2021designing} in their deployment in our daily lives.

\subsection{Humanoid Robots vs. Non-Humanoid Robots}
Many humanoid robots, such as NAO \cite{Monceaux2009demonstration} and Pepper \cite{pandey2018mass}, can employ readily available anthropomorphic features like body movement, facial expression or speech to exhibit emotions. While such physical embodiment may increase the recognition level of displayed emotions \cite{harris2011exploring,loffler2018multimodal}, a large class of social robots are tuned to be less- or non-anthropomorphic in order to match the functional requirements at their designated tasks \cite{gacsi2016humans,harris2011exploring,novikova2014design,song2017expressing}. For example, rescue robots are usually small and tank-like \cite{song2018designing}, and domestic vacuum robots like Roomba are likely to be puck-shaped and able to fit under couches \cite{harris2011exploring}. In addition to the utilitarian aspect, implementing emotions using the anthropomorphic features in humanoid robots not only can be expensive and technically complex \cite{gacsi2016humans,peng2020children,song2017expressing} but also needs to fulfill users' expectations on their level of anthropomorphism \cite{loffler2018multimodal,park2021should} and at the same time to minimise the feeling of creepiness known as the ``uncanny valley'' effect \cite{Herdel2021drone,park2021should}. Consequently, an increasing body of literature has focused on designing emotional expressions for non-humanoid robots to show affection. On the one hand, non-humanoid robots are anatomically unavailable to express emotions like humans; on the other, this reduces stereotypes of how emotions should be displayed and thus broadens the range of modalities across visual, auditory, and haptic channels that these robots may employ. In the context of this paper, we are interested in understanding what modalities have been used to encode emotions in non-humanoid robots and how well the resulting emotional expressions are identified and perceived by users.

\subsection{Current AV-Pedestrian Interaction}
With human drivers replaced by autonomous control systems, one important challenge for the social acceptance of AVs is to communicate their intent and awareness to nearby pedestrians and other vulnerable road users (VRUs) \cite{prattico2021comparing,rasouli2019autonomous}. A major direction in exploring effective communication strategies is the development of external human-machine interfaces (eHMIs) \cite{dey2020taming,prattico2021comparing,rasouli2019autonomous}. Current solutions to AV-pedestrian eHMIs are manifold. For example, vehicle-mounted LED lighting patterns have been utilised to indicate vehicle modes, the awareness of a nearby pedestrian, or the intention to yield or move \cite{lagstrom2016avip,Mahadevan2018communicating,rasouli2019autonomous}. Studies also investigated the use of eHMIs to convey messages by displaying pictograms \cite{urmson2015pedestrian} and texts\cite{Colley2021Investigating,clamann2017evaluation}. On-road projections have been investigated as a way to leverage traffic metaphors like crosswalks or stop signs \cite{Hesenius2018panic,locken2019automated,Nguyen2019designing}. In one implementation, the road infrastructure was updated to collect data from AVs and to convey the information to pedestrians via a smart road \cite{mairs2017umbrellium}. Some researchers also experimented with anthropomorphic features to restore current driver-pedestrian interaction patterns. An example for this is the implementation of moving eyes that follow the position of pedestrians at crosswalks \cite{chang2017eyes,locken2019automated}. Other studies tried a printed hand that waved to indicate yielding \cite{Mahadevan2018communicating} and a sign indicating a smile to inform pedestrians that it is safe to cross \cite{locken2019automated,prattico2021comparing}.

Those studies focused on the communication of intent and awareness through the use of operational cues, akin to the way traditional street signage does, which is designed to evoke immediate, and sometimes emotional, responses from users and further coordinate actions among interaction subjects. However, enabling AVs to express emotions as a communication strategy has not yet been addressed as a primary focus of research. This gap was also identified in a recently published design space for the external communication of AVs \cite{Colley2020Design}. The authors pointed out the need for ``affective messages'' (i.e. messages related to emotions) since such messages are highly important in interpersonal communication even if they do not necessarily carry meaning \cite{Colley2020Design}.

\subsection{Why to Ascribe Emotions to AVs}
The mass deployment of AVs in daily traffic environments could be hindered by their social acceptance related to not only technical aspects but societal factors as well \cite{prattico2021comparing}. Using eHMIs to communicate intent and awareness reduces public scepticism by improving pedestrians' understanding of the machine's decisions and maneuvers, and therefore fostering safe interaction. Yet, overcoming people's psychological barrier towards AVs' deployment is not easy, and various forms of discrimination against AVs are still continuously witnessed. For example, local residents harassed and attacked (e.g. thrown rocks at) Waymo's self-driving cars during a public trial because of feeling uncomfortable or scared around them \cite{waymo2018humans}. Volvo's driverless cars were reported to be ``easy prey'' on the road and were bullied by other drivers with slamming on brakes or aggressive driving to force them into submission \cite{volvo2016first}. Even being one of the most vulnerable road users, pedestrians were also found to take advantage of AVs' rule-abiding nature by crossing the road with impunity once they discovered that the cars were self-driving \cite{bazilinskyy2021driving,jayaraman2019pedestrian}. Though intelligent agents, AVs are often considered as mindless machines following programmed rules or even, more generally, a piece of ``creepy'' technology that breaks the status quo that people are comfortable with \cite{google2014revolution}.

Following findings from studies of social robots in other domains, a promising avenue to address some of these issues is to equip AVs with social traits, such as the ability to express emotions. This has the potential to shift people's perception of AVs as purely algorithmically driven agents towards intelligent social actors. Indeed, concepts for increasing AVs' sociability have surfaced in recent years. In 2014, Google announced a driverless car prototype that was intentionally designed like an adorable ``Marshmallow Bumper Bots'' with headlights like wide eyes and a front camera like a button nose, aiming to resemble a living being or a friend \cite{google2014revolution,google2014why}. Taking an even greater leap, in August 2021, Honda released a lively AV bot serving as both transportation and smart companion, conceptualised for the year 2040 \cite{honda2021niko}. It has a large frontal face with animated emotional facial expressions and fenders with covered wheels like pet animal legs, radiating ``the cute character of a playful puppy'' \cite{honda2021niko}. These efforts are striving to make AVs likable social agents and improve their acceptance by evoking people's empathy. In line with these concepts, increasing AVs' emotional expressiveness is likely to enrich their social characteristics \cite{hoffman2015design,Herdel2021drone,fong2003survey} and improve their acceptability \cite{cauchard2016emotion,Herdel2021drone}.

Apart from influencing perception and empathic understanding, emotion in HRI is known for its function to regulate and guide human decision-making and behaviour, biasing the interaction process away from negative or harmful results \cite{picard2003affective}. For example, a majority of in-car affective human-machine interfaces (HMIs) use expressive cues such as emotional music, ambient light, or empathic speech to regulate drivers into an emotionally balanced state and thus promote safe driving behaviour \cite{Braun2021affective}. Similarly, AVs expressing emotions through external affective cues may help regulate the traffic climate \cite{Sadeghian2020exploration}, especially when other road users disagree with AVs' decisions (e.g. road rage towards the AV's maneuver \cite{volvo2016first}). Close to the strategy of using emotions or affects, some existing AV-pedestrian eHMIs have attempted to convey courtesy through textual messages like ``Thank You'' and ``You're Welcome'' \cite{Colley2021Investigating}, and ``Please'' \cite{Lanzer2020designing}, facilitating the cooperation between AVs and pedestrians. Hence, along with adjusting people's preconceptions, AVs' expressiveness should help regulate and guide the interaction process and eventually contribute to their functioning. This kind of approach follows Picard's definition of affective computing, suggesting that it is not about making machines look ``more emotional'' but about making them more effective \cite{picard2003affective}.

The rich literature in affective robotics indicates that emotion encoding in AV-pedestrian interaction is possible. Indeed, humans have the innate tendency to attribute liveliness, emotions, intelligence, and other social characteristics to moving objects \cite{bretan2015emotionally,harris2011exploring,Hoggenmueller2020emotional,peng2020children,Tan2016happy,Tennent2018Character}. A close example to AVs are drones, also referred to as unmanned aerial vehicles (UAVs). These are small plane-like flying robots that show rich kinematics and functionalities \cite{cauchard2016emotion}. As personal drones are becoming more popular and ubiquitous in our daily lives \cite{cauchard2016emotion,Herdel2021drone}, emotion encoding in human-drone interaction (HDI) has also been studied in recent years \cite{cauchard2016emotion,Herdel2021drone,hieida2016action}, especially for the purpose of increasing their acceptability \cite{cauchard2016emotion, Herdel2021drone}. Through a systematic review of how drones and other non-humanoid social robots have displayed emotions to people surrounding them, we aim to investigate plausible emotions or affects as well as possible output modalities that could be considered for designing affective AV-pedestrian interfaces.

\section{Method}
To gain an understanding of current emotionally expressive non-humanoid robots, we reviewed relevant articles from 2011 to 2021 using a systematic search strategy.

\subsection{Search Strategy}
\subsubsection{Database Selection}
To identify the most relevant publishers, we first queried Google Scholar as it covers a broad search across various sources of publications. We used the Publish or Perish software (version 7) \cite{Harzing2007publish} to request Google Scholar in order to extract the search results into a CSV file, as Google Scholar does not provide the functionality to download the results as a whole. We searched for ``emotion'' AND ``robot'' from 2011 to 2021. We extracted the most relevant 1000 results as it was the maximum number of results available for retrieval. We then counted the number of results per publisher, resulting in four top publishers (IEEE = 291, Springer = 176, Elsevier = 72, ACM = 69) followed by Google Patents (31), MDPI (25), and SAGE (22) while others were below 20.

We then searched within the databases corresponding to the top four publishers, i.e. IEEE Xplore Digital Library, SpringerLink, ScienceDirect (for Elsevier), and ACM Digital Library, using the same query string and time frame as in the first step. We checked the number of results and then excluded SpringerLink as it yielded over 10,000 results under the ``Article'' and ``Chapter and Conference Paper'' content types, which would have made a subsequent review impossible to achieve. This left us with three databases for the detailed article search and analysis (ACM = 1,427, IEEE = 1,103, ScienceDirect = 3,329).

\subsubsection{Keyword Search Procedure}
To identify potential article candidates for the review, we conducted a keyword search within each of the selected databases. Three main keywords were used: ``emotion'', ``robot'', and ``non-humanoid''. We also included synonyms that are commonly used to describe a non-humanoid appearance for robots: ``non-anthropomorphic'', ``appearance-constrained'', and ``appearance constrained''. We combined these keywords using AND/OR operators and utilised the advanced search feature in each database. We selected a time frame of the last ten years because: (1) 87\% of the total results fell within the last ten years, and (2) we were interested in understanding recent trends in this growing discipline. The time frame for the final search was from 01/01/2011 to 16/07/2021. The search yielded a total of 225 results including research articles, posters, books, and other kinds of publications across the three databases (ACM = 77, IEEE = 77, ScienceDirect = 71).

\subsubsection{Article Selection}
We chose articles published in conference proceedings and journals that: (1) were written in the English language, (2) used a non-humanoid robot, (3) designed emotional expressions for the robot, and (4) evaluated the emotional expressions with empirical user studies and presented the evaluation results. In a first step, the lead author screened the articles. This process involved reading each article's title, abstract, and full-text to see if it met the selection criteria. In this process, we removed duplicated articles as some articles were published jointly by different publishers. We further excluded articles that proposed approaches for robots to sense or recognise users' emotions rather than express emotions, as well as articles that included a robot capable of displaying emotions but offered little information in how the emotional expressions were designed. The screening process resulted in the final collection of 25 articles that are included in our review. As a second step, two of the authors then reviewed and discussed the results, mapping them out on a digital whiteboard. No further changes were made to the article selection in that step.

\subsection{Research Questions}
\label{questions_review}
The review presented in this paper is based on the following research questions.
\begin{itemize}
\item What emotions are commonly expressed by non-humanoid robots?
\item How are the emotions displayed?
\item What measures are used to evaluate the emotional expressions?
\item What are the user perceptions of the emotional expressions?
\end{itemize}

\section{Review of Emotionally Expressive Non-Humanoid Robots}
This section presents the systematic review of the 25 articles in response to the research questions in Section~\ref{questions_review}.

\subsection{Overview}
\begin{figure}[h]	
\widefigure
\DIFdelbeginFL 
\DIFdelendFL \DIFaddbeginFL \includegraphics[width=\linewidth]{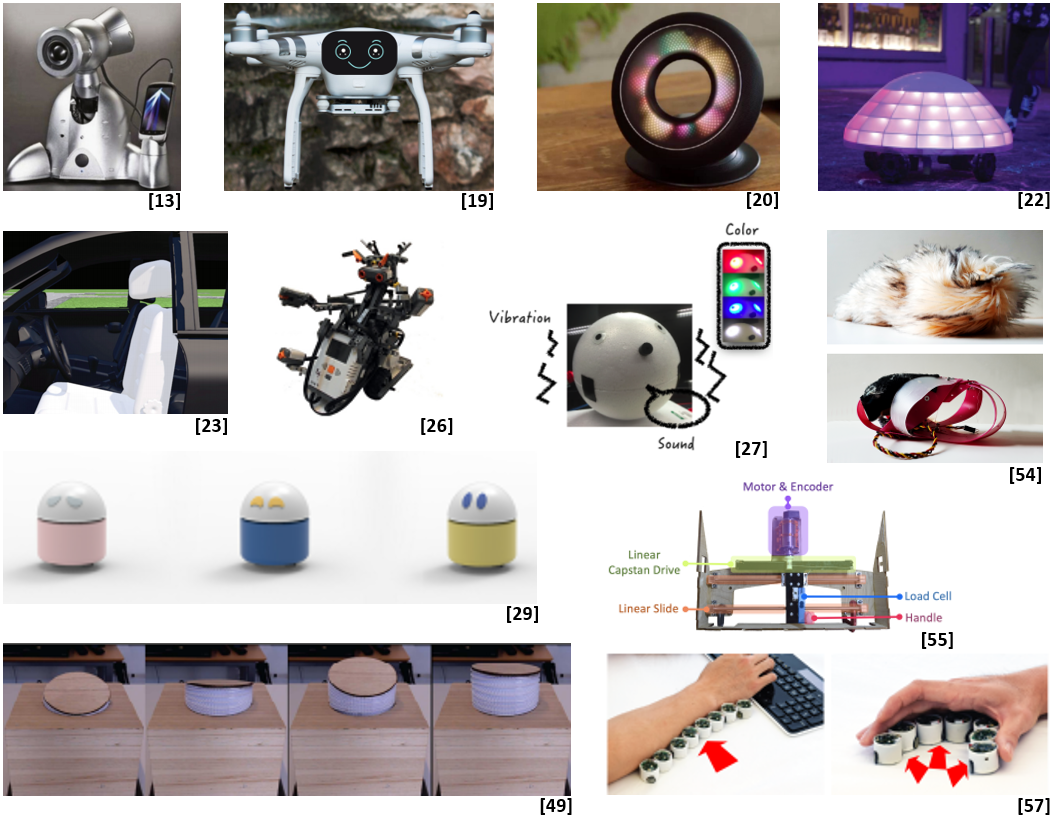}
\DIFaddendFL \caption{Examples of non-humanoid robots from the reviewed articles~\cite{bretan2015emotionally,Herdel2021drone,Whittaker2021designing,Hoggenmueller2020emotional,novikova2014design,song2017expressing,peng2020children,Tan2016happy,bucci2018happy,chase2019differences,kim2019swarm,Tennent2018Character}\hspace{0pt}
}
\label{fig:robots}
\end{figure}

An overview of the reviewed articles is provided in Table~\ref{tab:overview}. The non-humanoid robots used in these articles varied in morphologies and functionalities \DIFaddbegin \DIFadd{(Figure~\ref{fig:robots})}\DIFaddend . Eleven (44\%) articles adopted readily available robots developed in previous work \cite{bretan2015emotionally,gacsi2016humans,Hoggenmueller2020emotional,ritschel2019personalized,shi2018designing} or commercially available \cite{boccanfuso2015autonomously,cauchard2016emotion,frederiksen2020causality,hieida2016action,song2018designing,Whittaker2021designing}, while other thirteen (52\%) articles designed and prototyped robots for the specific purpose of investigating emotional expressions \cite{bucci2018happy,chase2019differences,frederiksen2020robots,harris2011exploring,hoffman2015design,kim2019swarm,loffler2018multimodal,peng2020children,sato2020affective,song2017expressing,Tan2016happy,Tennent2018Character,Herdel2021drone}. The remaining one article implemented the design from another robot using Lego robot parts \cite{novikova2014design}. Two articles developed animations instead of using a physical embodiment to simulate robots (a car seat in \cite{Tennent2018Character} and a drone in \cite{Herdel2021drone}). Regardless of whether the robots had a designated functionality per se, four robots had a perceivable utilitarian function during the evaluation of emotions, including two voice assistants \cite{shi2018designing,Whittaker2021designing}, one drone \cite{cauchard2016emotion}, and one car seat \cite{Tennent2018Character}. Another two robots served as companions during the studies reported in the articles \cite{boccanfuso2015autonomously,hoffman2015design}.

\end{paracol}
\begin{table}
\widetable
  \caption{Articles in the Review of Emotionally Expressive Non-Humanoid Robots}
  \label{tab:overview}
  \begin{tabular}{llp{4.2cm}p{4.5cm}p{3cm}}
    \toprule
    Year&Authors&Robot Prototype&Emotion Models&Output Modalities\\
    \midrule
    2011&Harris and Sharlin\cite{harris2011exploring}&The Stem&angry, happy, sad, etc.&movement\\[13pt]
    2014&Novikova and Watts\cite{novikova2014design}&a Lego robot based on a Phobot robot's design&Mehrabian's model&movement\\[13pt]
    2015&Boccanfuso et al.\cite{boccanfuso2015autonomously}&Sphero&angry, fearful, happy, sad&colour, movement, sound\\[13pt]
    2015&Bretan et al.\cite{bretan2015emotionally}&Shimi&Ekman's basic emotions,\newline Russell's circumplex model&movement\\[13pt]
    2015&Hoffman et al.\cite{hoffman2015design}&Kip1&calm, curious, scared&movement\\[13pt]
    2016&Cauchard et al.\cite{cauchard2016emotion}&AR.Drone 2.0 by Parrot&personalities from Walt Disney's Seven Dwarfs and Peyo's Smurfs&movement\\[13pt]
    2016&G{\'a}csi et al.\cite{gacsi2016humans}&PeopleBot&Ekman's basic emotions&movement, sound\\[13pt]
    2016&Hieida et al.\cite{hieida2016action}&``Rolling Spider'' Drone by Parrot&anger, joy, pleasure, and sadness from a Japanese idiom&movement\\[13pt]
    2016&Tan et al.\cite{Tan2016happy}&a shape-changing interface&Ekman's basic emotions, Mehrabian's model&movement\\[13pt]
    2017&Song and Yamada\cite{song2017expressing}&Maru&Russell's circumplex model&colour, movement, sound\\[13pt]
    2018&Bucci et al.\cite{bucci2018happy}&FlexiBit with a fur cover&emotional valence&haptics, movement\\[13pt]
    2018&L\"{o}ffler et al.\cite{loffler2018multimodal}&a simple, wheeled robot probe&Ekman's basic emotions&colour, movement, sound\\[13pt]
    2018&Shi et al.\cite{shi2018designing}&a smartphone-based voice assistant&Russell's circumplex model&facial expression, movement\\[13pt]
    2018&Song and Yamada\cite{song2018designing}&Roomba&Ekman's basic emotions, \newline Russell's circumplex model&colour, movement\\[13pt]
    2018&Tennent et al.\cite{Tennent2018Character}&3D animation for a robot car seat&aggressive, confident, cool,\newline excited, quirky&movement\\[13pt]
    2019&Chase and Follmer\cite{chase2019differences}&a device with a visible and graspable handle&Mehrabian's model&haptics, movement\\[13pt]
    2019&Kim and Follmer\cite{kim2019swarm}&SwarmHaptics&Ekman's basic emotions&haptics, movement\\[13pt]
    2019&Ritschel et al.\cite{ritschel2019personalized}&B\"{a}rBot&happiness, sadness, etc.&sound\\[13pt]
    2020&Frederiksen and Stoy\cite{frederiksen2020causality}&three Thymio II Robots&fear&movement, sound\\[13pt]
    2020&Frederiksen and Stoy\cite{frederiksen2020robots}&Affecta&remorse&colour, facial expression, movement, sound\\[13pt]
    2020&Hoggenmueller et al.\cite{Hoggenmueller2020emotional}&Woodie&Ekman's basic emotions&colour, movement\\[13pt]
    2020&Peng et al.\cite{peng2020children}&three small robot characters for a robot theatre&Ekman's basic emotions&facial expression, movement\\[13pt]
    2020&Sato et al.\cite{sato2020affective}&tabletop, wheeled haptic robots&Russell's circumplex model&haptics\\[13pt]
    2021&Herdel et al.\cite{Herdel2021drone}&animated drone with DJI Phantom 3 body&Ekman's basic emotions&facial expression\\[13pt]
    2021&Whittaker et al.\cite{Whittaker2021designing}&Olly&``Big Five'' personality theory&colour, movement, sound\\
  \bottomrule
\end{tabular}
\end{table}
\begin{paracol}{2}
\linenumbers
\switchcolumn

The robots were ascribed with a set of distinct expressive behaviours corresponding to specific emotions. These expressive behaviours were encoded through a variety of output modalities, supporting inferences of the robots' internal states and broadening the range of feasible cues applicable to designing affective interfaces. Twenty-three (92\%) articles investigated the encoding of multiple emotions and evaluated the effectiveness of one or more modalities in displaying the emotions, while the remaining two articles simply tested the perception and impact of a single emotion \cite{frederiksen2020causality,frederiksen2020robots}. To support a more interactive evaluation process, robots in eight (32\%) articles were able to express emotions based on the behaviours of participants \cite{boccanfuso2015autonomously,bretan2015emotionally,frederiksen2020robots,harris2011exploring,hoffman2015design,ritschel2019personalized,shi2018designing,Whittaker2021designing}.

The purpose of robotic emotion design in fifteen (60\%) articles was to explore the use of a modality or multiple modalities. Eight of these articles aimed to address the feasibility or effectiveness of the chosen modality/modalities in developing the emotional expressions \cite{harris2011exploring,bucci2018happy,Herdel2021drone,hieida2016action,loffler2018multimodal,song2018designing,song2017expressing,chase2019differences}, while the other seven of the fifteen articles focused on providing design strategies for using the modality/modalities to encode emotions \cite{novikova2014design,cauchard2016emotion,sato2020affective,Tennent2018Character,Tan2016happy,gacsi2016humans,kim2019swarm}. The remaining ten (40\%) articles had a main purpose other than evaluating modalities. Six of them examined the influence of emotional expressions on user cognition \cite{frederiksen2020causality,shi2018designing} or behaviour \cite{hoffman2015design,boccanfuso2015autonomously,frederiksen2020robots,peng2020children}. Two articles evaluated the efficacy of their proposed approach for the robot expressing emotions dynamically \cite{bretan2015emotionally,ritschel2019personalized}. As for the remaining two articles, one \cite{Whittaker2021designing} aimed to understand user preference for the robot's personality traits, and the other \cite{Hoggenmueller2020emotional} explored various contextual factors influencing user perception of the emotional expressions.

\subsection{Emotion Models}
To understand what emotions are commonly expressed by non-humanoid robots, this section presents the emotion models that guided the selection of emotions in the reviewed articles. We identified three emotion models, which are used to structure this section. Two streams of models are based on previous literature, namely {\itshape categorical models} and {\itshape dimensional models} \cite{bretan2015emotionally,Hoggenmueller2020emotional,novikova2014design,Tan2016happy}. We further added {\itshape emotional personas} as a third model since we found that some articles developed emotional personas for the robots and selected emotions in accordance with the personas to articulate corresponding personalities.

\subsubsection{Categorical Models}
Nineteen (76\%) articles encoded categorical emotions (e.g. happiness and sadness). The most employed categorical emotion model is Ekman's six basic emotions \cite{ekman1971constants} including anger, disgust, fear, happiness, sadness and surprise. These emotions are regarded as essential for human-human communication, easy to understand, and widely recognisable across cultural backgrounds \cite{bretan2015emotionally,Herdel2021drone,Hoggenmueller2020emotional,loffler2018multimodal}. Besides using a validated psychological model, two studies referred to cultural conventions. Hieida et al. derived anger, joy, pleasure, and sadness from a popular Japanese idiom, ki-do-ai-raku \cite{hieida2016action}, while Cauchard et al. \cite{cauchard2016emotion} chose emotional states (brave, dopey, grumpy, happy, sad, scared, shy, and sleepy) from personalities found in Walt Disney's Seven Dwarfs and Peyo's Smurfs.

\subsubsection{Dimensional Models}
Some studies argued that discrete emotions were unable to cover a comprehensive space of emotions since human emotions comprise not only basic emotions but also subtle variations within each category \cite{novikova2014design,sato2020affective}. For instance, joyous, content, and jubilant describe different levels of happiness \cite{bretan2015emotionally}. Consequently, dimensional approaches were utilised in nine (36\%) articles. This includes three articles \cite{bretan2015emotionally,Tan2016happy,song2018designing} that conducted multiple studies using both categorical emotion models and dimensional emotion models. Two mostly adopted dimensional models are Russell's circumplex model \cite{russell1980circumplex} and Mehrabian's model \cite{mehrabian1996pleasure}. Russell's circumplex model positions emotions on a two-dimensional space: valence and arousal, where valence refers to the positive or negative connotation of the emotion \cite{Hoggenmueller2020emotional}, and arousal means the intensity of the emotion. Mehrabian's model distributes emotions in a three-dimensional space: pleasure-arousal-dominance (PAD), also known as valence-arousal-dominance, in which the dominance dimension measures the controlling or submissive nature of the emotion \cite{Herdel2021drone}. For example, although both anger and fear are negative emotions, the former is perceived as dominant while the latter is submissive.

\subsubsection{Emotional Personas}
Two articles (8\%) used emotions to portray personalities for different personas \cite{cauchard2016emotion,Whittaker2021designing}. Instead of being presented with emotions directly, participants in these two articles were faced with stereotype emotional personas \cite{Whittaker2021designing} in which typical emotions were expressed and discerned. Cauchard et al. \cite{cauchard2016emotion} integrated selected emotional traits into three representative personas (adventurer, anti-social, and exhausted) for drones. For example, the behaviours of an adventurer drone showed a combination of happiness and bravery. In the work of Whittaker et al. \cite{Whittaker2021designing}, a voice-assisted home robot was assigned to three distinct personas (buddy, butler, and sidekick) which were derived based on the well-known ``Big Five'' personality traits \cite{goldberg1992development} and differed in perceived emotions via speech, intonation, colour, and movement when responding to people's commands.

\subsection{Output Modalities}
In order to answer how the emotions are displayed, this section demonstrates a variety of output modalities across visual, auditory, and haptic channels in the articles for creating affective interfaces for non-humanoid robots to manifest emotions. We classified the modalities into the sensory categories with reference to how these modalities were sensed by users during evaluation.

\subsubsection{Visual Modalities}
Twenty-three (92\%) articles utilised visual modalities including movement, colour, and facial expression. Many of the robots presented in the articles used a combination of visual modalities. 

The most employed modality was movement, which was found in twenty-two articles. Six articles used movement to encode emotions based on suggestions in prior work \cite{chase2019differences,boccanfuso2015autonomously,Hoggenmueller2020emotional,song2017expressing,Tan2016happy,Whittaker2021designing}. For instance, for a robot placed in a natural play scenario with young children, Boccanfuso et al. \cite{boccanfuso2015autonomously} used movements that were previously reported to be indicative of emotions in children. Tan et al. \cite{Tan2016happy} designed shape-changing movements by reviewing biological motion studies that had demonstrated relations between emotions and shape-changing parameters such as velocity and orientation. In addition to referencing previous findings directly, four articles created emotional movements based on validated knowledge systems including the Laban's movement framework \cite{sharma2013communicating} which was adopted in three articles \cite{hieida2016action,novikova2014design,Tennent2018Character} and the Interaction Vocabulary \cite{Lenz2013exploring} used in one article \cite{cauchard2016emotion}. Furthermore, metaphorical mappings were utilised in four articles \cite{gacsi2016humans,loffler2018multimodal,shi2018designing,song2018designing}. For example, L\"{o}ffler et al. \cite{loffler2018multimodal} used conceptual metaphors such as ``joy is up and active'' and ``anger is hot fluid in a container'' to develop movement patterns. Shi et al. \cite{shi2018designing} designed emotional movements for text boxes on a smart-phone based voice assistant using affective human body expressions. Besides prescribing movement--emotion mappings in advance, four articles explored how users specified the relationships between emotions and movements \cite{bucci2018happy,harris2011exploring,kim2019swarm,song2018designing}. Despite the non-humanoid form of the robots, six articles designed emotional body expressions \cite{bretan2015emotionally,frederiksen2020robots,gacsi2016humans,hieida2016action,hoffman2015design,novikova2014design}. Except for the body expressions for drones in \cite{hieida2016action} which still remained in the realm of mechanical movements, the other five articles followed anthropomorphic or zoomorphic behaviours such as stretching the ``neck'' for a lamp-like robot to show curiosity \cite{hoffman2015design} or borrowing emotional behavioural patterns from dogs \cite{gacsi2016humans}.

Coloured lights were used in seven articles. Three articles \cite{loffler2018multimodal,song2017expressing,song2018designing} corresponded one emotion to one colour, e.g. anger--red, sadness--blue, calm--white, etc. Among them, the two articles by Song and Yamada \cite{song2017expressing,song2018designing} generated the encoding based on prior research while the work of L\"{o}ffler et al. \cite{loffler2018multimodal} referred to conceptual metaphors like ``joy is light and warm'' and ``fear is darkness''. Additionally, four articles \cite{boccanfuso2015autonomously,frederiksen2020robots,Hoggenmueller2020emotional,Whittaker2021designing} combined multiple visual properties of coloured lights to reflect single emotions. For example, Hoggenmueller et al. \cite{Hoggenmueller2020emotional} developed colour patterns with animation effects, e.g. blur and fade green and yellow colours for showing disgust, and Boccanfuso et al. \cite{boccanfuso2015autonomously} used bright and colourful lights with high intensity to convey happiness. For generating the mappings between multiple colour properties and emotions, Whittaker et al. conducted an online user study to decide colour patterns for different emotional personas while the other three articles referenced prior work.

Four articles reported the design of facial expressions for conveying emotions \cite{frederiksen2020robots,Herdel2021drone,peng2020children,shi2018designing}. Herdel et al. \cite{Herdel2021drone} used the Facial Action Coding System (FACS) to establish associations between basic emotions and animated facial expressions on drones. Shi et al. \cite{shi2018designing} developed facial expressions for their smart-phone based voice assistant by first seeking well-received cartoon designs and then selecting the final expressions through an online survey. Instead of using multiple facial features, Peng et al. \cite{peng2020children} and Frederiksen and Stoy \cite{frederiksen2020causality} designed only eye patterns for displaying emotions. Peng et al. \cite{peng2020children} created eyes with various shapes and colours for their small theatre robots and identified the mappings with emotions using online questionnaires, while the animated eyes on a phone mounted on the robot in \cite{frederiksen2020causality} simply showed paying attention and conveyed an apologetic state in a scolding scenario.

\subsubsection{Auditory Modalities}
Eight (32\%) articles employed auditory modalities including non-linguistic utterances (NLUs), music, and vocalisations.

NLUs are usually defined as ``robot-like'' mechanical sounds (e.g. chirps, beeps, whirrs, etc.) \cite{read2016people} and are often used as affective cues in HCI \cite{read2016people,song2017expressing}. Five articles included NLUs for cueing emotions \cite{frederiksen2020causality,frederiksen2020robots,gacsi2016humans,loffler2018multimodal,song2017expressing}. For instance, mechanical sounds such as chirps and beeps varied in tones and rhythms were associated with different emotions in \cite{gacsi2016humans,loffler2018multimodal,song2017expressing}. In both articles by Frederiksen and Stoy \cite{frederiksen2020causality,frederiksen2020robots}, NLUs were used to show single emotions (e.g. alerting audio signals to express fear \cite{frederiksen2020causality} and augmented naturally occurring sounds of the robot to convey remorse \cite{frederiksen2020robots}).

Two articles utilised music that was validated to evoke emotions before \cite{boccanfuso2015autonomously,ritschel2019personalized}. Ritschel et al. \cite{ritschel2019personalized} used previously proposed melodies to show emotions and intentions and personalised the timbre dynamically according to user preferences. Boccanfuso et al. \cite{boccanfuso2015autonomously} produced a set of synthesised music to enhance the robot's emotional expressions in the play environment with young children (e.g. a happy state was conveyed with a piece of music with frequent and smooth changes in a moderate to high pitch).

Vocalisations were found in two articles \cite{boccanfuso2015autonomously,Whittaker2021designing}. Whittaker et al. \cite{Whittaker2021designing} implemented humanoid speech and intonation to articulate three distinct personas in the voice-assisted home robot, while Boccanfuso et al. \cite{boccanfuso2015autonomously} added non-linguistic child vocalisations, such as giggle and crying, simply to augment the main sound cue (i.e. music).

\subsubsection{Haptic Modalities}
Haptic modalities, including haptic movements and textures, were used in four (16\%) articles. In one of those articles, the robot was covered in a naturalistic fur to mimic furry animals and invite user touch only to assist in the evaluation of the primary modality (i.e. breathing behaviours) \cite{bucci2018happy}. The other three articles used haptics as main cues \cite{chase2019differences,kim2019swarm,sato2020affective}. In the work of Sato et al. \cite{sato2020affective}, they investigated how users mapped combinations of haptic movements (e.g. tap rapidly/slowly) and textures (e.g. aluminium, clay, etc.) to a list of discrete emotions. Chase and Follmer \cite{chase2019differences} combined haptic movements with visual movements to test the perceived pleasure-arousal-dominance (PAD) for properties like stiffness and jitter. Kim and Follmer \cite{kim2019swarm} assessed perceived PAD in a swarm of small haptic devices by changing parameters including the number of robots, force types, frequency, and amplitude.

\subsection{Evaluation Measures}
All of the reviewed articles included user evaluations to assess the quality and impact of the emotional expressions. This section discusses the evaluation measures in terms of use scenarios, experimental tasks, and evaluated aspects.

\subsubsection{Use Scenarios}
Eleven (44\%) articles created use scenarios for robots displaying emotions in the evaluation. Three of the articles embedded the emotional expressions into the robots' tasks \cite{cauchard2016emotion,shi2018designing, Whittaker2021designing}. For instance, the drone in \cite{cauchard2016emotion} displayed emotional profiles during different flying tasks. The two voice assistants in \cite{shi2018designing,Whittaker2021designing} conveyed emotional states during tasks activated by users' spoken commands, e.g. setting a reminder \cite{shi2018designing} and playing a music \cite{Whittaker2021designing}. Six articles created scenarios specifically for the emotional expressions \cite{frederiksen2020causality,frederiksen2020robots,peng2020children,Tennent2018Character,novikova2014design,hoffman2015design}. The car seat in \cite{Tennent2018Character} showed expressive movements when greeting its driver. Peng et al. designed plots for the robot actors in the robot theatre to show contextual emotions \cite{peng2020children}. The conversation companion robot in \cite{hoffman2015design} responded to humans' vocalics during a conflict conversation between couples. The robot in \cite{frederiksen2020robots} was placed in a scolding scenario in order to convey remorse. Two articles intentionally designed triggers for the emotions in order to help with users' understanding, such as a scenario where the robot showed positive emotions after finishing its task successfully \cite{novikova2014design} and a high volume explosion sound to evoke the robot's fearful reactions \cite{frederiksen2020causality}. Using a more natural context, Boccanfuso et al. placed the robot into an unstructured play environment with young children for eliciting their affective responses \cite{boccanfuso2015autonomously}. Though the robot in \cite{bretan2015emotionally} expressed emotions without contexts in the first two studies, it responded to users' speech in the third one. Apart from these eleven articles, the robots' emotional expressions in other articles were devoid of any use case, presumably with the purpose of evaluating the design without bias \cite{harris2011exploring,Hoggenmueller2020emotional}.

\subsubsection{Experimental Tasks}
In seven articles (28\%), participants viewed images or watched pre-recorded or simulated videos where the robot showcased emotional expressions \cite{bretan2015emotionally,gacsi2016humans,Herdel2021drone,hieida2016action,novikova2014design,peng2020children,Tennent2018Character}. Specifically, the videos in \cite{Tennent2018Character} and \cite{Herdel2021drone} were simulated animations. Nineteen (76\%) articles presented participants with physically embodied robots, including the article by Bretan et al. \cite{bretan2015emotionally} in which participants were faced with either the physical robot or images/videos. Six of those articles asked participants to only watch the robots displaying emotions \cite{cauchard2016emotion,frederiksen2020causality,Hoggenmueller2020emotional,loffler2018multimodal,song2017expressing,song2018designing}, whereas thirteen of the articles let participants interact with the robots \cite{boccanfuso2015autonomously,bretan2015emotionally,bucci2018happy,chase2019differences,frederiksen2020robots,harris2011exploring,hoffman2015design,kim2019swarm,ritschel2019personalized,sato2020affective,shi2018designing,Tan2016happy,Whittaker2021designing}. The contents of the interactions varied across articles. For example, participants were instructed to touch or play with the robots in order to experience the emotional expressions \cite{boccanfuso2015autonomously,bucci2018happy,chase2019differences,harris2011exploring,kim2019swarm,sato2020affective,Tan2016happy}. In the two articles with voice-assisted robots \cite{shi2018designing,Whittaker2021designing}, participants received emotions from the robot while completing specified tasks using the robot. Furthermore, some robots were able to adjust their emotional responses according to participants' behaviours through either pre-programmed mechanisms \cite{boccanfuso2015autonomously,bretan2015emotionally,frederiksen2020robots,hoffman2015design,ritschel2019personalized} or a Wizard of Oz procedure \cite{harris2011exploring}.

\subsubsection{Evaluated Aspects}
Twelve articles (48\%) gauged the recognition level of the emotional expressions, that is, the rate that an emotion was correctly recognised through its expression. In nine articles, participants matched presented expressions with a list of emotion names \cite{cauchard2016emotion,gacsi2016humans,Herdel2021drone,loffler2018multimodal,novikova2014design,sato2020affective,song2017expressing,Tan2016happy,Tennent2018Character}, whereas the other three articles asked participants to rate how much they thought the presented expressions matched the prescribed emotions on scales \cite{bretan2015emotionally,Hoggenmueller2020emotional,ritschel2019personalized}.

Other characteristics of the emotional expressions were also measured. Five articles used the Self-Assessment Mannequin (SAM) scale \cite{bradley1994measuring} to measure the perceived valence, arousal, and dominance of the expressions \cite{chase2019differences,Herdel2021drone,kim2019swarm,song2018designing,Tan2016happy}. Several articles evaluated valence \cite{bucci2018happy,frederiksen2020causality,hieida2016action}, arousal (or intensity) \cite{cauchard2016emotion,frederiksen2020causality,Herdel2021drone,hieida2016action}, and dominance \cite{frederiksen2020causality} independently. Two articles \cite{harris2011exploring,hieida2016action} asked participants to rate each emotional expression on a scale of pairs of opposite adjectives like ``tired vs. energetic'' and ``hasty vs. leisurely''. Other traits such as emotional and cognitive engagement \cite{shi2018designing}, perceived urgency \cite{kim2019swarm}, and the impact of emotions on surroundings \cite{frederiksen2020robots,Herdel2021drone} were also measured. Additionally, some articles assessed the level of anthropomorphism of their robot. For instance, two articles employed HRI metrics for gauging the perceived anthropomorphism, likeability and safety of the robots \cite{chase2019differences,kim2019swarm}. Similarly, social human character traits like friendliness, cooperativeness, sociability, etc., as well as participants' comfort level with the robot were assessed in one of the articles \cite{hoffman2015design}.

Three articles analysed participants' behaviours during their interaction with the robots using video coding \cite{boccanfuso2015autonomously,hoffman2015design} or direct observation \cite{harris2011exploring}. Through the analysis of video coding data, Boccanfuso et al. \cite{boccanfuso2015autonomously} characterised different play patterns and affective responses of young children, and Hoffman et al. \cite{hoffman2015design} collected verbal references to the robot during simulated couple conflicts. In the direct observation procedure in \cite{harris2011exploring}, think-aloud was used for participants to express feelings and thoughts during their interaction with the moving robot. Moreover, nine articles presented and discussed participants' comments on their experience with the emotionally expressive robot, either from interviews or open-ended questions in questionnaires \cite{cauchard2016emotion,bucci2018happy,harris2011exploring,Herdel2021drone,hoffman2015design,Hoggenmueller2020emotional,peng2020children,ritschel2019personalized,Whittaker2021designing}.

\subsection{User Perceptions}
\label{perception}
To discuss user perceptions of the emotional expressions in the reviewed articles, we summarise common findings reported including the recognition level of emotions and other important aspects associated with user perceptions.

\subsubsection{Recognition of Emotional Expressions}
\label{perception_recognition}
Basic emotions that are relatively obvious and universal \cite{cauchard2016emotion,Tan2016happy} were best recognised by participants, including happiness \cite{cauchard2016emotion,gacsi2016humans,Herdel2021drone,Hoggenmueller2020emotional,novikova2014design,peng2020children}, sadness \cite{bretan2015emotionally,Herdel2021drone,Hoggenmueller2020emotional,ritschel2019personalized}, and anger \cite{Hoggenmueller2020emotional,peng2020children}. However, sadness was at the same time found to be the least recognisable emotion in three articles \cite{gacsi2016humans,novikova2014design,peng2020children}. Two other negatively valenced emotions, fear \cite{bretan2015emotionally,Herdel2021drone} and disgust \cite{Herdel2021drone,Tan2016happy}, were also the most difficult to identify. This is in line with research in psychology showing that some emotions are more easily recognised than others \cite{Herdel2021drone} and that negative emotions tend to be recognised slower \cite{Herdel2021drone} and are less consistently interpreted correctly \cite{Tennent2018Character} compared to positive emotions. Furthermore, emotions like surprise, coolness, and affirmation that are believed to be more abstract and need more contexts to interpret \cite{Hoggenmueller2020emotional,kim2019swarm} were also rated low in terms of recognition \cite{Hoggenmueller2020emotional,ritschel2019personalized,Tennent2018Character}.

When combining emotions with modalities, we found that the interpretation of emotional expressions was mostly the intuitive comprehension of the emotion--expression relationship. For example, in most studies where movement was used, participants were likely to associate high speed or high frequency with emotions with high arousal such as excitement or anger \cite{cauchard2016emotion,harris2011exploring,Hoggenmueller2020emotional,kim2019swarm,novikova2014design,song2017expressing,Tan2016happy}, while a low speed or an avoidance behaviour were usually interpreted as less intense emotions like sadness or fear \cite{Hoggenmueller2020emotional,novikova2014design,song2018designing}. Similarly, participants were able to understand intuitive, conventional mappings for other modalities. For instance, a falling sound or slow tempo were usually interpreted as conveying sadness \cite{loffler2018multimodal,ritschel2019personalized,song2017expressing}, and bright and fast-changing colours were commonly associated with joy \cite{Hoggenmueller2020emotional,loffler2018multimodal,Whittaker2021designing}.

\subsubsection{Sociability}
Many articles reported that participants attributed liveliness, internal states, and sociability to the emotionally expressive robots. For example, participants in one article \cite{bucci2018happy} referred to the emotional furry robot as having ``a rich inner life'' and reminding them of pets. In the storytelling section in another article \cite{peng2020children}, children interpreted motivations, intentions and emotions from the performance of the robots in the robot theatre. In \cite{hoffman2015design}, the companion robot was perceived as friendly, warm, and capable of forming social bonds and attachments. The drone with facial expressions in \cite{Herdel2021drone} was described as an agent with autonomy, consciousness, and cognitive and behavioural abilities.

Nevertheless, disengagement was found when robots displayed certain emotional expressions. Harris and Sharlin \cite{harris2011exploring} reported that nearly half of their participants showed boredom when presented with slow and repetitive movements. For the voice-assisted home robot in \cite{Whittaker2021designing}, the ``sidekick'' persona which had a low amplitude voice and used slow movements also failed to engage some participants emotionally. Besides, Boccanfuso et al. \cite{boccanfuso2015autonomously} suggested that negative emotions that elicited frustration and annoyance might cause disengagement in children. In the investigation of emotional and cognitive engagement of the emotionally expressive voice assistant, Shi et al. \cite{shi2018designing} found that emotions with positive valence and high arousal might help robots to establish emotional connections with humans. 

\subsubsection{Contexts}
Context was regarded as an important factor influencing users' interpretation of emotional expressions. In \cite{novikova2014design}, the recognition rate of emotions improved dramatically when emotions were displayed within an appropriate context compared to displayed alone. Tan et al. \cite{Tan2016happy} also suggested that adding a use scenario could help users identify and disambiguate emotions. In order to interpret relatively more abstract emotions such as fear and surprise, participants in \cite{kim2019swarm} tended to combine the haptic expressions with other factors such as motion paths and the contact locations of the haptic stimuli to obtain more contextual information. Hoggenmueller et al. \cite{Hoggenmueller2020emotional} discussed a range of contextual aspects that impacted users' comprehension, including spatiotemporal context, interactional context, as well as contexts related to users' background.

Additionally, participants were found to create narratives for the emotional expressions in several articles. For instance, participants in \cite{Herdel2021drone} tended to develop stories to make sense of the emotions, such as speculating about the cause of fear and surprise. Children who watched the theatre play performed by multiple robots in \cite{peng2020children} generated conjectures about robots' social relationships according to the display of emotions. After analysing participants' verbal descriptions of the expressive behaviours of the robot, Bucci et al. \cite{bucci2018happy} concluded that narratives made about the motivation and situation of the robot could heavily influence the perception of emotional expressions.

\section{Considerations}
The reviewed studies show rich evidence for designing affective interfaces for non-humanoid robots to communicate emotions. This serves as a foundation and offers guidance for adding an emotional dimension to AV-pedestrian interfaces. In this regard, we aim to provide preliminary considerations around core elements for designing affective AV-pedestrian interfaces. First, we draw on findings from the review to provide a set of considerations for designing emotional expressions for AVs as social robots. Then, based on both the review and current AV-pedestrian communication strategies, we present a set of considerations that take a broader range of factors into account for building affective AV-pedestrian interfaces.

\subsection{Considerations for Designing Emotional Expressions}
Based on findings from the review of emotional expressions of non-humanoid robots, we propose five considerations for imbuing AVs with emotions with regard to what emotions to communicate, and how to communicate emotions.
\\
\\
\emph{Include Basic Emotions:}
More than half of the reviewed articles employed basic emotions, which were mostly derived from Ekman's six emotions \cite{ekman1971constants}. Particularly, the user evaluation in these articles showed that happiness, sadness, and anger were most easily recognised \cite{cauchard2016emotion,bretan2015emotionally,gacsi2016humans,Herdel2021drone,Hoggenmueller2020emotional,novikova2014design,peng2020children,ritschel2019personalized}. In general, such emotions are regarded as easy to understand, recognisable across different cultural backgrounds, and important for intuitive human-robot interaction \cite{bretan2015emotionally,Herdel2021drone,Hoggenmueller2020emotional,loffler2018multimodal}. Hence, we suggest that basic emotions or basic emotion models should be considered when deciding what emotions to attribute to AVs.
\\
\\
\emph{Use Negative Emotions for a Reason:}
In some articles where negative emotions were triggered for a reason, these emotions demonstrated important contributions in affecting user behaviours, e.g. defused conflict situations when showing fear \cite{hoffman2015design} or remorse \cite{frederiksen2020causality}, and evoked sympathetic behaviours when displaying sadness \cite{boccanfuso2015autonomously}. Indeed, robot emotions can cause humans to mirror the emotional state of the robot \cite{frederiksen2020causality,Herdel2021drone,Hoggenmueller2020emotional} or to reflect on their own behaviours \cite{frederiksen2020robots,Herdel2021drone,hoffman2015design}. However, when negative emotions were displayed without reasons, users speculated about the cause of the emotions \cite{Herdel2021drone}, involuntarily interpreted them as positive valenced \cite{Hoggenmueller2020emotional}, or even sometimes concerned about their own safety under aggressive emotions \cite{chase2019differences,harris2011exploring,Tennent2018Character}. Therefore, providing reasons for the display of negative emotions can be essential for the intended user interpretation and the subsequent influence on user behaviours.
\\
\\
\emph{Provide Contexts for Abstract Emotions:}
Some reviewed articles showed that emotions such as surprise, disgust, coolness, and affirmation, though some included in Ekman's six basic emotions, are more abstract in connotation than those universally recognisable emotions (e.g. happiness and sadness) and thus need more contexts for users to interpret them as intended \cite{Hoggenmueller2020emotional,kim2019swarm,ritschel2019personalized,Tennent2018Character}. When displayed without context, these emotions can seem to be ambiguous \cite{Tan2016happy}, and the interpretation can be more greatly biased by the user's cultural background and previous experiences \cite{Hoggenmueller2020emotional}. Hence, the expected user perception of abstract emotions in AVs can hardly be isolated from an appropriate context, a context relating but not limited to the task that the AV is performing \cite{novikova2014design}, the immediate surroundings \cite{Hoggenmueller2020emotional}, and cultural norms \cite{park2021should}.
\\
\\
\emph{Combine Multiple Modalities:}
Several studies compared the recognition rate of emotions between using single and multiple modalities \cite{loffler2018multimodal,song2017expressing,song2018designing}. Results showed that people recognised the multi-modal expressions more easily and were more confident in their judgement. Indeed, when multiple modalities are combined together to convey a certain emotion, they tend to serve as an ``amplifier'' to each other \cite{song2018designing} and reassure people of their interpretation \cite{loffler2018multimodal,song2017expressing}. Even in some studies where only a single modality was tested, some participants still referred to partial cues other than the primary modality to support their inferences \cite{Herdel2021drone,kim2019swarm}. Hence, using multiple modalities can be beneficial to clarifying the emotional state of the AV and help increase users' confidence in making fast and safe decisions accordingly.
\\
\\
\emph{Employ Intuitive Encoding:}
The review showed that emotions were best interpreted when the encoding followed a conventional and intuitive assignment of expression--emotion relationships, such as using colourful lights or uplifting music for happiness \cite{boccanfuso2015autonomously,Hoggenmueller2020emotional,loffler2018multimodal,ritschel2019personalized,Whittaker2021designing} and slow movements or dark colours for sadness \cite{Hoggenmueller2020emotional,loffler2018multimodal,song2017expressing,song2018designing}. Nonetheless, such ``intuitive'' mappings can be culture-specific \cite{peng2020children,frederiksen2020robots,loffler2018multimodal,song2018designing}. For instance, some studies argue cultural differences in mapping colour to emotion \cite{peng2020children}. Besides, there are also conceptual models or universal associations that are very little dependent on culture or can be found in many languages \cite{loffler2018multimodal,peng2020children}. Overall, employing encoding that is intuitive to users is important in AV-pedestrian interaction, as such interaction requires immediate decisions in dynamic traffic situations. Similar concerns can be found in existing AV-pedestrian eHMIs. For example, a participant in \cite{Mahadevan2018communicating} commented that during the crossing scenario, they had to frequently look at a sheet which specified the mapping between multiple LED colours and vehicle states, and they further pointed out that it was unrealistic to bring a sheet in real life. Thus, it should be carefully considered to use encoding rules that might require a learning process.

\subsection{Considerations for Building Affective AV-Pedestrian Interfaces}
Drawing on the review and taking into account current AV-pedestrian communication strategies, this section presents a set of considerations around a broader range of factors that may contribute to designing affective AV-pedestrian interfaces.
\\
\\
\emph{Align with AV's Primary Functionality:}
AVs' primary function (i.e. transportation) is likely to influence people's interpretation of their social traits during the interaction. In the review, only four robots had a perceivable primary function during user evaluation \cite{shi2018designing,Whittaker2021designing,cauchard2016emotion,Tennent2018Character}. However, as shown in \cite{harris2011exploring} and \cite{Hoggenmueller2020emotional}, when interacting with a robot, people are likely to speculate about the functionality or ``purpose'' of the robot and interpret its emotions accordingly. This suggests that people would expect the emotional expressions of a robot to be coherent with its functionality. In the example of a voice-assisted home robot \cite{Whittaker2021designing}, people preferred the smart helper to show conscientiousness and agreeableness through its expressive cues. Therefore, the affective interface should account for its interplay with AVs’ major utilitarian purpose, i.e. serving as a secondary function \cite{Hoggenmueller2020emotional} and facilitating the operation of the primary one.
\\
\\
\emph{Understand Pedestrian's Expectations:}
Pedestrians' expectations of AVs' emotions can differ from passengers sitting inside the vehicle or other road users. Various interaction contexts have been found in AV-pedestrian interaction, such as interaction contents, road types, other vehicles or road users, etc \cite{tran2021review}. If AVs are to use emotions to support the interaction with pedestrians, it is important to consider the legitimacy of the emotions in those interaction contexts from the perspective of pedestrians, that is, to understand what emotions pedestrians expect to see. Similar considerations have been reported in human-drone interaction (HDI). For instance, in the design of emotional drones, Cauchard et al. \cite{cauchard2016emotion} left aside emotions that did not seem to be applicable to HDI, such as disgust. Herdel et al. \cite{Herdel2021drone} also stated the concern about whether users would envision certain emotions (e.g. fear) to appear in drones. Besides, emotional patterns in existing driver-pedestrian interaction may provide a vital reference for conjecturing pedestrians' expectations of AVs' emotions, as it is essentially the drivers' role that the autonomous systems take over when interacting with pedestrians socially.
\\
\\
\emph{Refer to Existing eHMIs:}
Although there is no empirical evidence yet for which modality is effective for an affective AV-pedestrian interface, efforts made in current eHMIs for communicating AVs' intent and awareness offer various solutions, such as vehicle-mounted displays \cite{chang2017eyes,clamann2017evaluation,lagstrom2016avip,Mahadevan2018communicating,urmson2015pedestrian}, on-road projections \cite{Hesenius2018panic,locken2019automated,Nguyen2019designing}, smart-road interfaces \cite{mairs2017umbrellium}, and wearable devices \cite{dey2020taming}. These interfaces provide a range of feasible modalities across visual, auditory, and haptic channels to support the potential communication of emotions. Nevertheless, many previous studies show that regardless of the presence of eHMIs, pedestrians still greatly rely on changes in the movement of AVs (e.g. speed) \cite{clamann2017evaluation,epke2021see,lee2021road,pillai2017virtual,rasouli2019autonomous}. Thus, the affective interface may prefer movement cues to encode emotions (e.g. movement-related ``gestures'' \cite{dey2020taming}), or use emotions to amplify the intention of movements (e.g. display a happy face to show friendliness when the vehicle is yielding). Overall, designers should refer to existing AV-pedestrian eHMIs and their corresponding findings to understand the usability of different types of interfaces when designing for the affective dimension.
\\
\\
\emph{Make It a Reciprocal Process:}
Current eHMIs for conveying intent and awareness to pedestrians are mostly designed in a proactive manner. For example, in a crossing scenario, most visual displays and on-road projections provide information about the deceleration progress of the vehicle \cite{clamann2017evaluation,Hesenius2018panic,prattico2021comparing} or inform pedestrians when it is safe to cross \cite{Nguyen2019designing,urmson2015pedestrian}. However, it is also important for AVs to be responsive to pedestrians' intentions \cite{epke2021see}. A recent study by Epke et al. \cite{epke2021see} used human gestures and eHMIs to form a bi-directional communication between pedestrians and AVs. The study found that participants preferred the case where an approaching AV acted (i.e. displaying ``I SEE YOU'' or yielding) in accordance with pedestrians' hand gestures. More importantly, the communication of emotions is itself a reciprocal process in which emotional responses can be evoked between the two subjects \cite{Herdel2021drone}. Hence, AVs should have the capability to express emotions not only proactively but also in response to the behaviours of pedestrians or contingencies in their surroundings.
\\
\\
\emph{Account for Contextual Factors:}
As suggested in previous sections, the display of emotion should be appropriate and plausible within the specific context. However, a number of contextual factors can condition pedestrians' interpretation of the emotional expression, such as pedestrians' own background, the environment where the AV is situated in, and the way the AV interacts with pedestrians \cite{fischer2019emotion,Hoggenmueller2020emotional,novikova2014design,pillai2017virtual}. For instance, cultural norms can influence how users perceive the emotional expression \cite{fischer2019emotion} and also the appropriateness of the situation in which the emotion is expressed \cite{park2021should}. In the deployment of an emotional self-moving robot in a university campus \cite{Hoggenmueller2020emotional}, Hoggenmueller et al. found that participants generally perceived the robot as in a happy state regardless of what emotion the robot was displaying, due to the peaceful environment and the robot's high luminosity. Moreover, weather and road conditions like rainy and poor lighting can impact the effectiveness of the AV's interface \cite{pillai2017virtual,rasouli2019autonomous}. Hence, the design of affective AV-pedestrian interfaces should account for various contextual factors, including cultural, environmental, and interactional aspects.
\\
\\
\emph{Consider Evaluation Environments:}
Most AV-pedestrian interaction solutions are evaluated in lab-based simulated environments since building fully functional prototypes on actual AVs can be expensive, complicated, and sometimes dependent on traffic regulations \cite{tran2021review}. On the one hand, simulation-based prototypes such as virtual reality (VR) simulations are more feasible and flexible than physical ones deployed in the wild; on the other, they are still weaker in terms of ecological validity \cite{Albastaki2020augmenting}. Furthermore, VR-based evaluations can be divided into immersive VR (i.e. use wearable equipment and immerse users in the virtual world) and non-immersive VR (i.e. users experience the virtual world via a computer screen) \cite{Albastaki2020augmenting}, in which the former provides a better presence of the AV and the latter is less confined by hardware and allows for remote user testing. Nevertheless, studies in affective robotics show that the proximity and presence of the robot can impact people's perceptions of its social intentions \cite{bretan2015emotionally,Tennent2018Character}. Therefore, evaluation environments should be considered and compared, particularly in the context of both AV-pedestrian interaction and affective robotic design.

\section{Conclusion}
In this paper, we investigated the potential role of affective interfaces for AVs to communicate emotions to pedestrians. To understand possible solutions to this problem, we systematically reviewed 25 articles about the emotional expressions of non-humanoid robots in the past ten years. The review summarised core aspects including emotion models, output modalities, evaluation measures, as well as user perceptions. Building on findings from the review and taking into account current AV-pedestrian communication strategies, we proposed five considerations for designing emotional expressions for AVs and six considerations regarding a broader range of factors contributing to building affective AV-pedestrian interfaces. Our findings provide a foundation for incorporating an affective dimension into AV-pedestrian eHMIs to increase the social acceptance of AVs and highlight avenues for investigating these opportunities in future research.

\authorcontributions{Conceptualization, Y.W., L.H. and M.T.; methodology, Y.W., L.H. and M.T.; formal analysis, Y.W.; writing---original draft preparation, Y.W.; writing---review and editing, Y.W., L.H. and M.T.; visualization, Y.W.; supervision, L.H. and M.T. All authors have read and agreed to the published version of the manuscript.}

\funding{This research was funded by the Australian Research Council through grant number DP200102604 Trust and Safety in Autonomous Mobility Systems: A Human-centred Approach.}

\institutionalreview{Not applicable.}

\informedconsent{Not applicable.}



\conflictsofinterest{The authors declare no conflict of interest.}

\end{paracol}
\reftitle{References}


\externalbibliography{yes}
\bibliography{references}


%


\end{document}